\documentclass[a4paper,fleqn,usenatbib]{mnras}

\usepackage{newtxtext,newtxmath}

\usepackage[T1]{fontenc}
\usepackage{ae,aecompl}

\usepackage{graphicx}	
\usepackage{amsmath}	
\usepackage{amssymb}	

\title[Accretion and photodesorption of CO vs angle]{Accretion and photodesorption of CO ice as a function of the incident angle of deposition}

\author[Gonz\'{a}lez D\'{i}az et al.]{
C. Gonz\'{a}lez D\'{i}az,$^{1}$\thanks{E-mail: 
	\href{mailto:cgonzalez@cab.inta-csic.es}{cgonzalez@cab.inta-csic.es}; \href{maito:munozcg@cab.inta-csic.es}{munozcg@cab.inta-csic.es}
    \href{maito:asperchen@phy.ncu.edu.tw}{asperchen@phy.ncu.edu.tw}}
H. Carrascosa de Lucas,$^{1}$
S. Aparicio,$^{2}$
G. M. Mu\~noz Caro, $^{1}$\footnotemark[1]\newauthor
N.-E. Sie,$^{3}$
L.-C. Hsiao,$^{3}$
and Y.-J. Chen$^{3}$\footnotemark[1]\\
$^{1}$Centro de Astrobiolog\'{\i}a (CSIC-INTA), Ctra. de Ajalvir, km 4, Torrej\'on de Ardoz, 28850 Madrid, Spain\\
$^{2}$Instituto de Tecnolog\'{\i}as F\'{\i}sicas y de la Informaci\'on, Leonardo Torres Quevedo, ITEFI (CSIC), c/ Serrano 144, 28006 Madrid, Spain\\
$^{3}$Department of Physics, National Central University, Jhongli District, Taoyuan City 32054, Taiwan}

\date{Accepted XXX. Received YYY; in original form ZZZ}

\pubyear{2019}

\begin{document}
\label{firstpage}
\pagerange{\pageref{firstpage}--\pageref{lastpage}}
\maketitle

\begin{abstract} 
	
Non-thermal desorption of inter- and circum-stellar ice mantles on dust grains, in particular ultraviolet photon-induced desorption, has gained importance in recent years. These processes may account for the observed gas phase abundances of molecules like CO toward cold interstellar clouds. Ice mantle growth results from gas molecules impinging on the dust from all directions and incidence angles. Nevertheless, the effect of the incident angle for deposition on ice photo-desorption rate has not been studied. This work explores the impact on the accretion and photodesorption rates of the incidence angle of CO gas molecules with the cold surface during deposition of a CO ice layer. Infrared spectroscopy monitored CO ice upon deposition at different angles, ultraviolet-irradiation, and subsequent warm-up. Vacuum-ultraviolet spectroscopy and a Ni-mesh measured the emission of the ultraviolet lamp. Molecules ejected from the ice to the gas during irradiation or warm-up were characterized by a quadrupole mass spectrometer. The photodesorption rate of CO ice deposited at 11 K and different incident angles was rather stable between 0 and 45$^{\circ}$. A maximum in the CO photodesorption rate appeared around 70$^{\circ}$-incidence deposition angle. The same deposition angle leads to the maximum surface area of water ice. Although this study of the surface area could not be performed for CO ice, the similar angle dependence in the photodesorption and the ice surface area suggests that they are closely related. Further evidence for a dependence of CO ice morphology on deposition angle is provided by thermal desorption of CO ice experiments.	
	
\end{abstract}



\section{Introduction}
\label{Introduction}
Dust grains observed in the infrared toward regions in space protected from the external ultraviolet (UV) radiation field, such as dense interstellar cloud interiors and circumstellar regions near the equatorial plane of the disk, reach temperatures below 20 K. With the exception of H$_2$, all molecules accrete on the dust under these conditions. Carbon monoxide, CO, is one of the most volatile molecules detected in ice mantles. The weak permanent dipole of CO allows its detection in the ice, by means of infrared spectroscopy, toward lines of sight where the light emitted by an infrared source is absorbed by ice-covered dust particles \citep{Lacy1984}. A few works report the gas-to-ice relative abundances of CO in dense clouds \citep[and ref. therein]{Pineda_2010}. Uncertainties in the solid-to-gas ratio of CO will affect the estimated CO/H$_2$ ratio often used to calculate the total interstellar gas mass, at least on the local scales of the cloud \citep{1Williams1985}. 

It is therefore essential to understand the interaction between the gas and solid phases. The formation of CO occurs efficiently in the gas phase. At temperatures below 26.6 K, CO was found to accrete onto the cold surface inside an ultra-high vacuum set-up. The accretion rate of CO in this experiment was constant at different accretion temperatures between 26.6 K and 8 K \citep{Cazaux_2017}. The low temperatures in dark cloud interiors should therefore allow a rapid CO ice growth and forbid thermal desorption of the ice. Several studies were conducted to study the thermal desorption of astrophysical CO ice analogs \citep[e.g.,][]{Collings2004,MunozCaro2010,MartinDomenech2014,Fayolle2016,Cazaux_2017,Luna2017,Luna2018}.  Because a significant fraction of the CO molecules in cold dense clouds is present in the gas phase, non-thermal desorption processes are proposed to explain these observations. In the laboratory, simulations of the desorption of CO ice molecules driven by energetic photons \citep{Oberg2007,Oberg2009,MunozCaro2010,Fayolle2011,Chen_2014} and cosmic rays \citep{SeperueloDuarte2010} were performed. UV-photodesorption of CO is very efficient compared to other ice molecules. 

Only the most energetic component of the UV radiation field, with energies between 11.1--13.6 eV, allow the direct dissociation of CO molecules. In most experiments, the cut-off imposed by  the MgF$_2$ window acting as the interface between the UV lamp and the vacuum  chamber is at 10.9 eV. This reduces the formation rate of photoproducts  and, therefore, the infrared CO band can be used to estimate its photodesorption rate. Other common ice components tend to dissociate delivering a more efficient  photochemistry than CO and a considerably lower photodesorption efficiency \citep[and ref. therein]{MartinDomenech2018}.  But experiments using a monochromatic UV beam of photon energy above the  CO dissociation energy also lead to an efficient photodesorption of CO molecules \citep{Bertin_2013}.

A drop in the CO photodesorption rate was observed for ice thickness below 5 $\pm$ 1 monolayers (ML). The measured quantum yield of CO photodesorption is higher than unity, expressed in number of photodesorbed molecules per absorbed photon in the top 5 ML. This suggests that energy transfer between molecules in CO ice is an important mechanism. In this process, a photoexcited molecule becomes electronically and vibrationally excited. This energy is transfered to other neighbor molecules.
When a molecule on the ice surface is sufficiently excited, a conversion to translational energy occurs, allowing this molecule to break the bond/(s) with its neighboring molecules and desorb to the gas phase. In two-component ices, e.g., CO:N$_2$, the photon energy absorbed by the CO molecules is also transferred to N$_2$ molecules and vice-versa, leading to the photodesorption of both species \citep{Bertin_2013}.   

\cite{Fayolle2011} reported a clear correlation between the UV absorption cross section and the photodesorption rate at different monochromatic wavelengths. This work was extended by \cite{Chen_2014}, using continuum-emission microwave discharge hydrogen lamp (MDHL) of T and F-types with different emission spectra, to study the effect on the CO photodesorption rate and the formation of CO$_2$. A recent paper \citep{MunozCaro2016} studied the effect of CO ice deposition temperature in the photodesorption rate, previously reported by \cite{Oberg2009}. One of the conclusions in this work was that the linear drop in the photodesorption rate for increasing deposition temperature was not related to a transition from amorphous to crystalline ice. The observed frequency shifts in the infrared and UV bands of CO ice occurred at deposition temperatures above 20 K, and was later attributed to the presence of Wannier-Mott excitons in UV-irradiated CO ice \citep{Chen2017}.

In this work, we propose to study the effect on the CO accretion and photodesorption rates of a different initial parameter: the incidence angle of the deposition tube with respect to the cold substrate. The astrophysical motivation of this work rests on the fact that gas molecules impinge on dust grains in all directions, spanning across the full range of incidence angles with the dust surface. 

\section{Experimental protocol}
\label{Experimental}
The CO ice irradiation and warm-up experiments employed the InterStellar Astrochemistry Chamber (ISAC), an ultra-high-vacuum (UHV) set-up with a base pressure of $4.0 \times 10^{-11}$ mbar; for a full description of ISAC, see \cite{MunozCaro2010}. A capillary tube of 1 mm internal diameter is connected through a needle valve to the gas line. This capilar is pointing to the substrate KBr window at a distance of about 3 cm. During deposition of the ice layer on the cold substrate window at 11 K, the needle valve is opened until the CO pressure in the UHV chamber reaches 2 $\times$ 10$^7$ mbar. Ice deposition at different angles is achieved by rotating the head of the cryostat to a given angle using an electrical engine controlled by a home-made software. The series of experiments reported in this paper followed this procedure. The estimated error in the deposition angle is less than 1$^{\circ}$. Because no radiation shield was employed in these experiments to allow ice deposition at glancing angles, the deposition temperature of 11 K was higher than the typical 8 K in the ISAC set-up. The sample holder housing the infrared window is fixed at the tip of the cold finger of a closed-cycle He cryostat, which was rotated to deposit the ice at different angles while the deposition tubes remain at a fixed position. Fourier-transform infrared spectroscopy (FTIR) of the ice in transmittance was performed using a Bruker VERTEX 70 at a spectral resolution of 1--2 cm$^{-1}$,  at normal incidence angle with respect to the sample substrate. The ice column density in molecules cm$^{-2}$, $N$, was estimated using the formula\\
\begin{equation}
\begin{centering}
N = \frac{1}{A} \int_{band} {\tau_{\nu}d\nu}
\end{centering}
\label{N}
,\end{equation}
where ${\tau}_{\nu}$ is the optical depth of the infrared band in absorption, $d\nu$ the wavenumber differential in cm$^{-1}$, and $A$ the band strength in cm molecule$^{-1}$. Here the integrated absorbance is equal to 0.43 $\times$ $\tau$, where $\tau$ is the integrated optical depth of the band. For CO ice, we used a band strength value of $A$(CO)=1.1 $\times$ 10$^{-17}$ cm molecule$^{-1}$ \citep{Jiang1975}, and for CO$_2$ formed upon irradiation of CO ice we used $A$(CO$_2$)=7.6 $\times$ 10$^{-17}$ cm molecule$^{-1}$ \citep{Yamada1964}. 

The CO ice deposition rate estimated by infrared spectroscopy, corresponding to a chamber pressure of $P$ = 2 $\times$ 10$^{-7}$ mbar, ranged from (2.5--0.8) $\times$ 10$^{14}$ molecules cm$^{-2}$ s$^{-1}$ depending on the angle of deposition. This dependence of the deposition rate with the deposition angle will be discussed in Sect. ~\ref{astro_implications}. Between 220-280 ML ($2 - 2.8 \times10^{17}$ molecules cm$^{-2}$) of CO ice were deposited at 11 K at different angles of incidence with respect to the substrate normal.  Frequency shifts in the CO ice band that result from different incident angles are much smaller than the typical spectral resolution of 1-2 cm$^{-1}$. Yet, they could be observed. This is explained in \cite{MunozCaro2016}. In short, the capability to measure a peak shift is not only related to the working resolution, since it is also related to the peak wavenumber accuracy of the FTIR instrument, better than 0.005 cm$^{-1}$ at 2000 cm$^{-1}$, i.e. near the position of the solid CO peak. 

\begin{figure}
	\centering
	\includegraphics[width=\hsize]{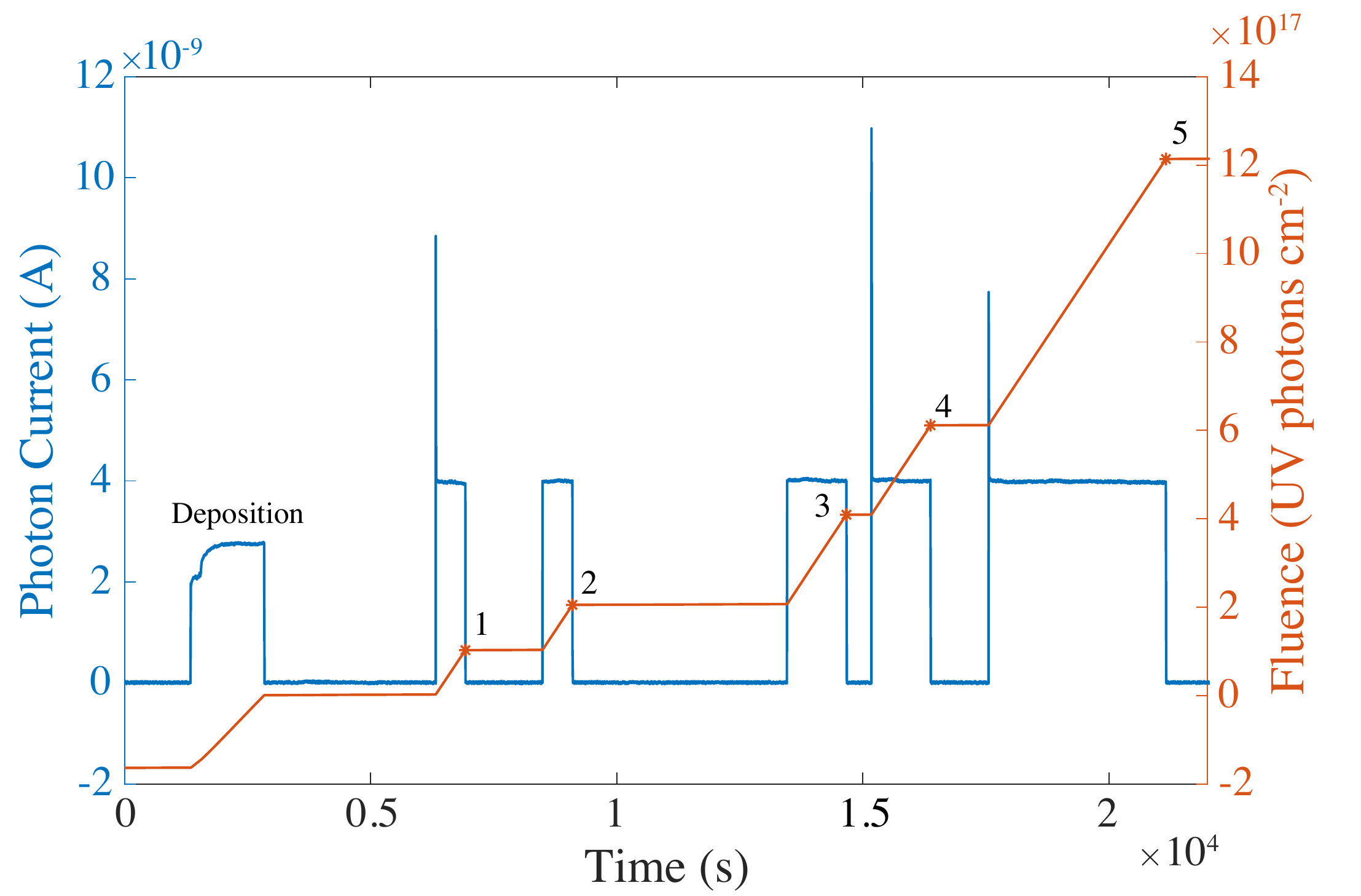}
	\caption{Photon current and fluence provided by the MDHL during various irradiation intervals.}
	\label{Fig_fluence}
\end{figure}

CO ice samples were UV-irradiated using a microwave-stimulated hydrogen flow discharge lamp (MDHL). Its emission spectrum in the vacuum-ultraviolet (VUV) was reported in \cite{CruzDiaz2014} and discussed in \cite{Chen_2014}. A quartz light guide was placed between the MDHL and the sample substrate to direct the UV photons to the ice. For this work, in addition to the McPherson 0.2 meter focal length VUV monochromator (Model 234/302) equipped with a photomultiplier tube (PMT) detector to monitor the VUV emission spectrum of the MDHL, we also measured the total value of the VUV flux using a callibrated Ni-mesh installed at the end of the light guide, i.e. about 3 cm away from the sample substrate. A typical photon flux of 2.5 $\times$ 10$^{14}$ photons cm$^{-2}$ s$^{-1}$  at the sample position is derived from the photon electron current in the quantum factor calibrated Ni-mesh. The fluence was then estimated multiplying the VUV flux values measured by the Ni-mesh by the irradiation time. Fig.~\ref{Fig_fluence} shows a sequence of irradiation intervals, visualized as an increase in the photon current (in Amperes) after the MDHL is turned on and a drop when the MDHL is turned off. Prior to irradiation, the MDHL is turned on for 20 min to stabilize the VUV flux, also displayed in this figure, which remains quite constant afterwards. The estimated fluence values, dots connected with solid lines in red, are displayed on scale.

Temperature-programmed desorption of the irradiated ice was performed at a constant heating ramp of 1 K min$^{-1}$ and monitored by FTIR spectroscopy. The desorbing molecules were detected using a quadrupole mass spectrometer (QMS) equipped with a Channeltron detector (Pfeiffer Vacuum, Prisma QMS 200).

\section{Experimental results}
\label{Simulations}
\begin{figure} 
	\centering
	\includegraphics[width=\hsize]{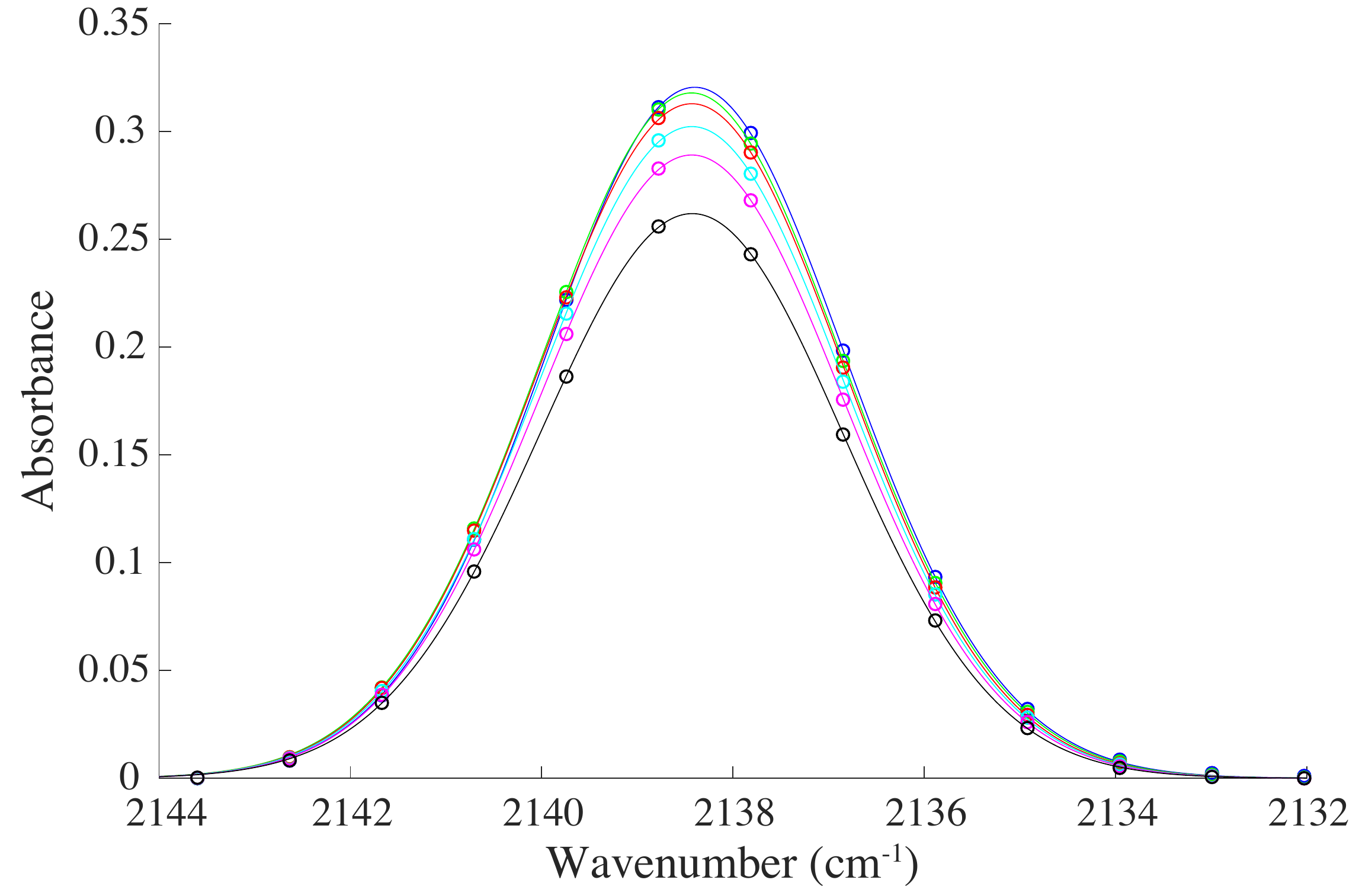}
	\caption{Infrared spectra of CO ice deposited at 11 K and an incident angle of 30 $^{\circ}$.}
	\label{Fig_IR_band}
\end{figure}

As an example of a typical experiment in this article, Fig.~\ref{Fig_IR_band} shows the CO ice spectra for a 30$^{\circ}$  deposition angle and a substrate temperature of 11 K, collected at different VUV irradiation times, see inlet. Open circles correspond to the data points collected at a spectral resolution of 2 cm$^{-1}$. A two-Gaussian deconvolution was applied to these datapoints, solid line, which provides an excellent fit. The central band position was around 2138 cm$^{-1}$, as expected for solid CO. The integrated absorbances were estimated from these fits and entered in Eq.~\ref{N} to calculate the column densities. The column density of the deposited CO ice in this experiment was $2.71 \times10^{17}$ molecules cm$^{-2}$, i.e. 271 monolayers (ML), where 1 ML is defined as  $1 \times10^{15}$ molecules cm$^{-2}$. 

\begin{figure}
	\centering
	\includegraphics[width=\hsize]{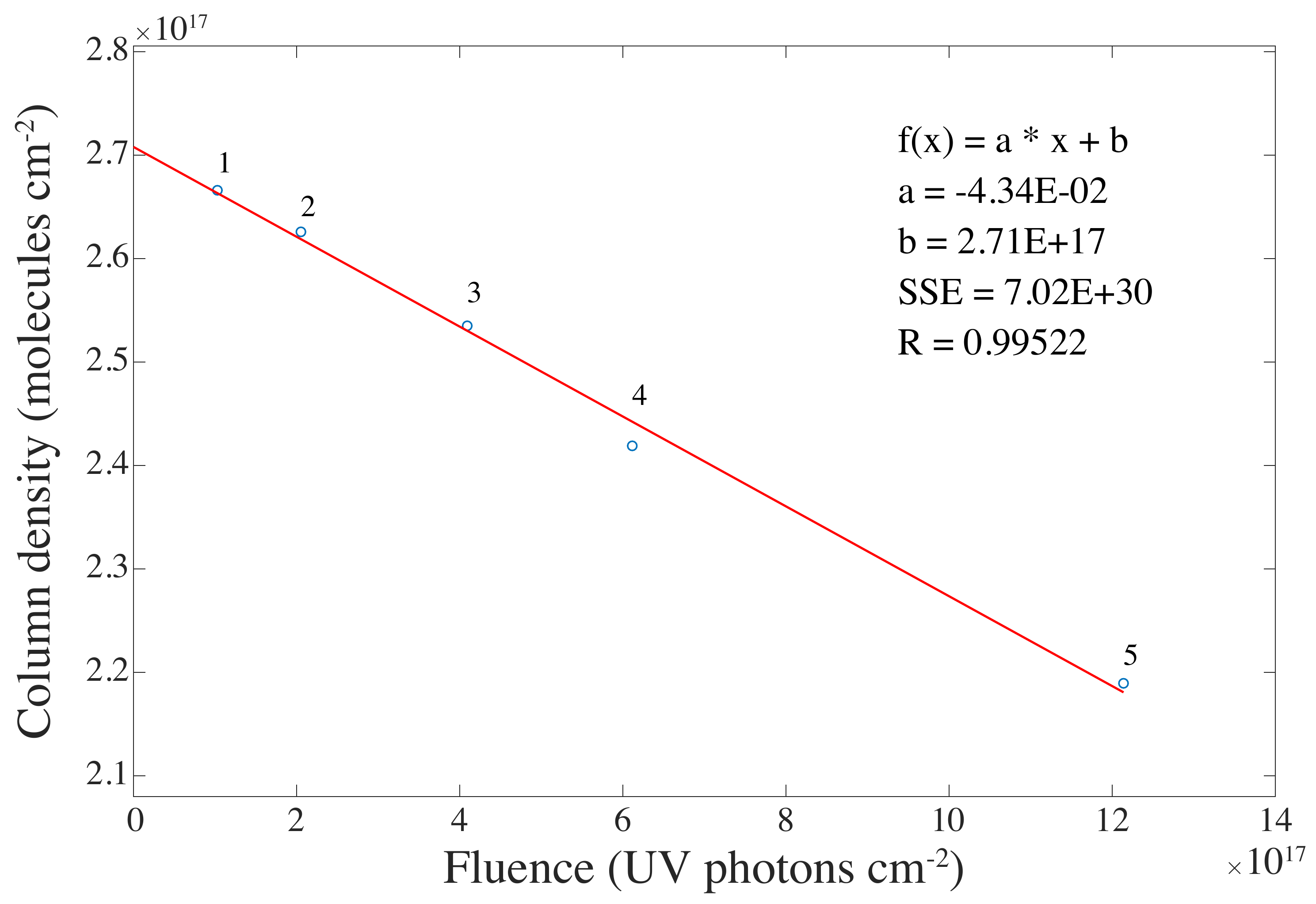}
	\caption{Linear decrease of the CO ice column density obtained from band integration in Fig.\ref{Fig_IR_ position}  as a function of VUV fluence. In this experiment, CO ice was deposited at 11 K and 30$^{\circ}$ angle of deposition. The slope of the linear fit corresponds to a photodesorption rate of 4.34 10$^{-2}$ molecules / UV photon.}
	\label{Fig_IR_N}
\end{figure}

The ice column densities during irradiation as a function of VUV fluence are represented in Fig.~\ref{Fig_IR_N}, open circles, and fitted linearly to estimate the slope. The goodness of fit parameters included in Fig.~\ref{Fig_IR_N} are the sum of squares error (SSE) that accounts for the deviation from the data, and the regression factor R that ranges from 0 to 1. 
A similar linear fit, not shown, is obtained if the band intensities are plotted, i.e. the height of the Gaussian fits, which means that the full width at half maximum (FWHM) remains constant during irradiation at the working resolution of 2 cm$^{-1}$. 
Because photoproduct formation in thin CO ice irradiation experiments is negligible (e.g., Mu\~noz Caro et al. 2010), the decrease of the band during irradiation results mainly from photodesorption, and the slope of the linear fit is equivalent to the photodesorption rate. In this particular experiment, the estimated photodesorption rate was 4.34 $\times$ 10$^{-2}$ molecules per incident UV photon.   

\begin{figure}
	\centering
	\includegraphics[width=\hsize]{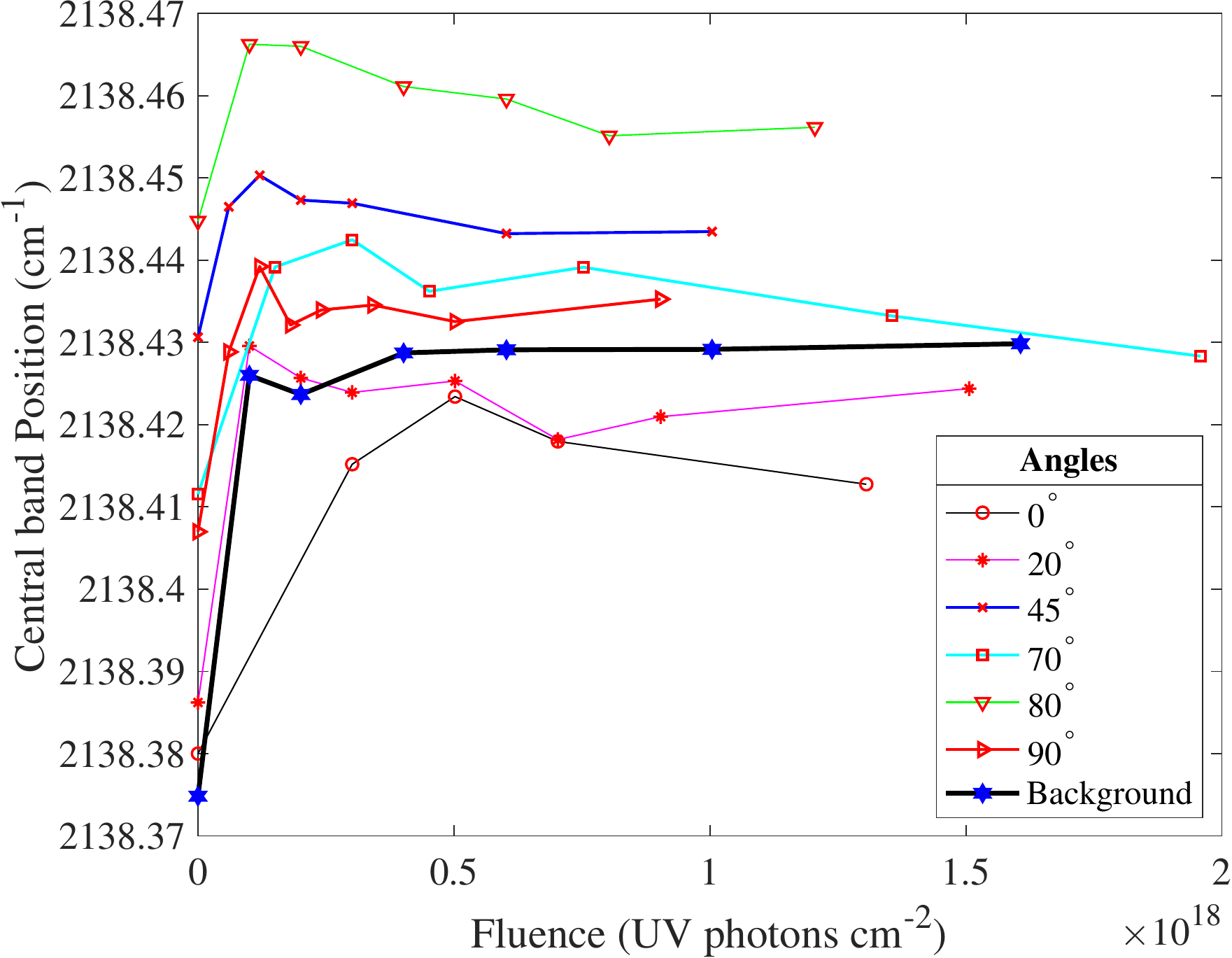}
	\caption{Central band positions of CO ice, deposited at different angles, as a function of UV fluence}
	\label{Fig_IR_ position}
\end{figure}

We used the same Gaussian fits to plot the central band position of CO ice during irradiation at normal position of the VUV beam with respect to the sample substrate, as shown in Fig.~\ref{Fig_IR_ position}. The observed shifts were certainly much smaller than the working spectral resolution of 1-2 cm$^{-1}$ but larger than the wavenumber accuracy of 0.005 cm$^{-1}$. In addition, they were reproducible and can be related to the optical ice properties, as explained in Sect.~\ref{Experimental} and \cite{MunozCaro2016}. The common trend for all the deposition angles is that there is a shift in the central band position after the first irradiation interval, which then tends to remain unchanged for larger VUV fluences within the experimental uncertainties. Because the formation of photoproducts was negligible, this observation is compatible with a change in the physical ice properties, such as compaction or a variation in the molecular disorder of the ice toward a more stable configuration. These studies were conducted for UV irradiated and ion processed water ice, which infrared spectrum is more sensitive to structural changes than the infrared band of CO \citep{Leto2003,Dartois2015}.
 
\begin{figure}
	\centering
	\includegraphics[width=\hsize]{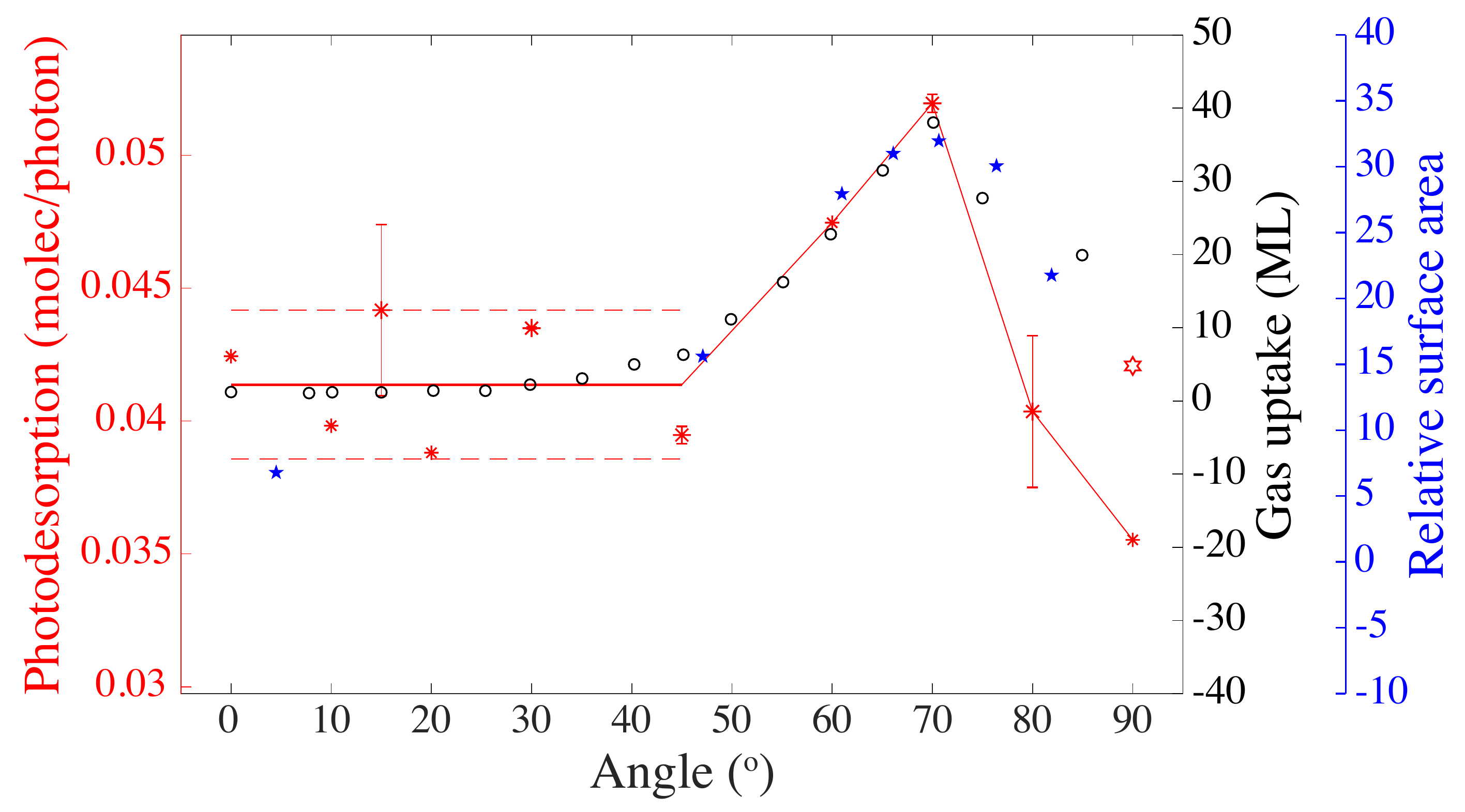}
	\caption[]{Photodesorption rate, in molecules per incident photon in the ice  as a function of CO ice-deposition angle (red asterisks). Red empty star corresponds to the CO photodesorption rate for the background deposition experiment. Black empty dots represent the Ar uptake by amorphous water ice, adapted from ref. \cite{Dohnalek2003}. Blue stars represent the effective surface area relative to the substrate area for thin film with cylinder shaped hypercolums, adapted from ref. \cite{Suzuki2001}.}
	\label{Fig_pico70}
\end{figure}

The photodesorption rate was estimated from the slope of the linear fit, as shown in Fig.~\ref{Fig_IR_N}, for all the experiments performed at different deposition angles. The result  is shown in Fig.~\ref{Fig_pico70} where the red asterisks are the photodesorption rates at each deposition angle. Error bars correspond to the standard deviation obtained from 2 or 3 repeated experiments at the same angle of deposition. Dashed red lines are the error limits in the 0 -- 45$^{\circ}$ range. The solid red lines are to guide the eye. The red star represents the photodesorption rate for a background deposition experiment, where the deposition tube was not in the line of sight of the cold substrate. Fig.~\ref{Fig_pico70} displays a clear peak in the photodesorption rate at larger deposition angles with a maximum near 70$^{\circ}$.   
This remarkable behavior of the photodesorption rate in experiments performed at different deposition angles,  and the comparison to the gas uptake by amorphous water ice in \cite{Dohnalek2003} (black empty dots) and the effective surface area computed by \cite{Suzuki2001} (blue stars) will be discussed in Sect.~\ref{astro_implications}.

\begin{figure}
	\centering
	\includegraphics[width=\hsize]{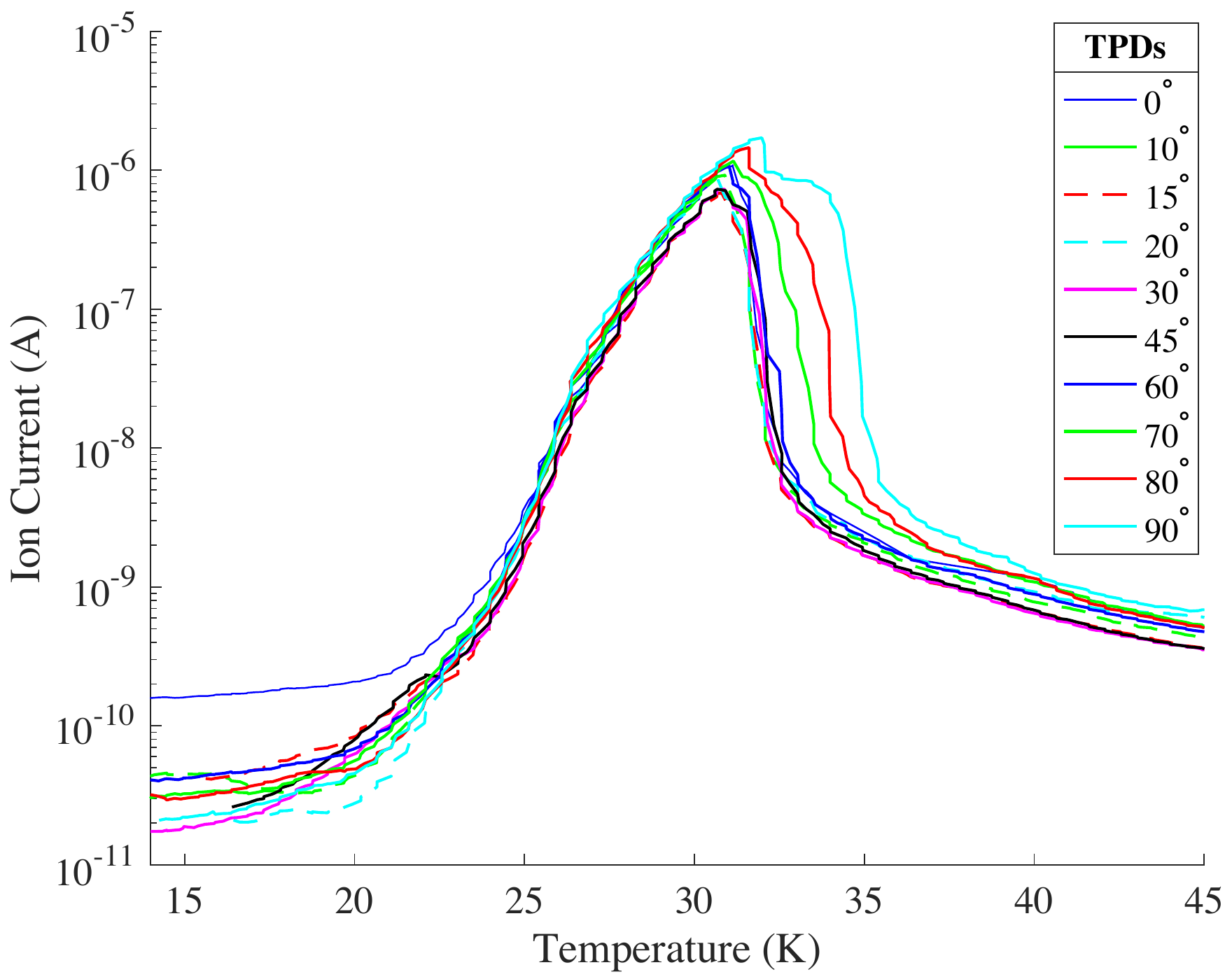}
	\caption{TPD of CO ice deposited and irradiated at 11 K for different angles of deposition.}
	\label{Fig_TPD}
\end{figure}

After the irradiation sequence was completed, temperature programmed desorption (TPD) experiments of the remaining CO ice was performed with a linear heating rate of  1 K min$^{-1}$. A quadrupole mass spectrometer (QMS) was used to monitor the volatile species in the gas phase. Figure~\ref{Fig_TPD} shows the TPD curves of these ices. The peak in the desorption displays a maximum near 30 K, in agreement with our previous experiments \citep[e.g.,][]{MunozCaro2010,Cazaux_2017}. Above 60$^{\circ}$ deposition angle, there is a clear shift in the right wing of the TPD peak toward higher temperatures, which demands an explanation. First, this shift suggests that the expectedly low diffusion of molecules deposited around 11 K in our experiments, is not sufficient to remove the morphological differences of ices deposited at different angles. Indeed, if a significant diffusion process would occur during CO ice deposition, VUV irradiation, or warm-up, the TPD curves would show no dependence on the angle of deposition. Second, the shifts observed clearly above 60$^{\circ}$ indicate that the ice morphology is significantly different for CO ice samples deposited at large grazing angles. This effect might be related to the peak in the photodesorption rate, with a maximum at 70${^{\circ}}$, see Figure~\ref{Fig_pico70}. In conclusion, according to our photodesorption and TPD results, there seems to be a pronounced change in the CO ice morphology for samples deposited at incidence angles $\geq$ 60${^{\circ}}$. This observation will be discussed in Sect.~\ref{astro_implications}.

\section{Astrophysical implications and conclusions}
\label{astro_implications}
The profile of the infrared band attributed to CO ice, observed toward young stellar objects and dense interstellar clouds, was fitted with 3 components: a weakly polar environment such as pure CO ice, CO mixed with weakly polar or apolar species, or CO mixed with polar species. Our work is more directly applicable  to the CO band component observed at 2139.7 cm$^{-1}$ that is matched by pure CO ice measured in the laboratory \citep[e.g.,][]{Pontoppidan2003,Penteado2015,Zamirri2018}. This layer of CO-rich ice is likely deposited on top of the already existing ice mantle, which is composed of water and other species detected in the ice. The deposition temperature of CO under ultra-high vacuum is 26.6 K in our experiments \citep{Cazaux_2017}, which corresponds to an astrophysical temperature near 20 K due to the longer timescales in space \citep{Collings2004,MartinDomenech2014}. Depending on its accretion temperature onto dust grains, in the 20--6 K range, CO-rich ice layers will display a more or less amorphous structure \citep{Cazaux_2017}. 
In this work, we exploited the ability to deposit the ice at different incident angles of the cold substrate relative to the deposition tube, to explore the impact on CO accretion and photodesorption. A variable incident angle is often used as a tool to grow films under different controlled conditions. This method is known as oblique angle deposition (OAD), \citep[see, e.g.,][]{Suzuki2001,Flaherty2012,BARRANCO2016}. Despite its popularity in Surface Science, only rarely was the effect of incident angle on ice morphology studied in a systematic manner. The majority of experimental simulations of ice growth were performed at normal incidence of the gas molecules on the surface, which account for head-on sticking, while some works employ a background or non-directed deposition, where either no deposition tube is used, or this tube is not pointing directly to the cold substrate. 

\cite{Dohnalek2003} deposited water vapour onto a cold substrate with variable incident angles to form amorphous water ice. The water ice density decreased with increasing deposition angle. At large incidence, they measured an ice porosity value of up to 80\%. In addition, the surface area of the water films was characterized using the adsorption of weakly bound gases (N$_2$, Ar, and CH$_4$) on the water ice surface. From angles of 0$^{\circ}$ to 30$^{\circ}$ the amount of the adsorbed gas was rather constant. For larger angles the amount of adsorbed gas increased, reached a maximum around 70$^{\circ}$, and then decreased gradually as the deposition angle approached 90$^{\circ}$. 

\begin{figure}
	\centering
	\includegraphics[width=\hsize]{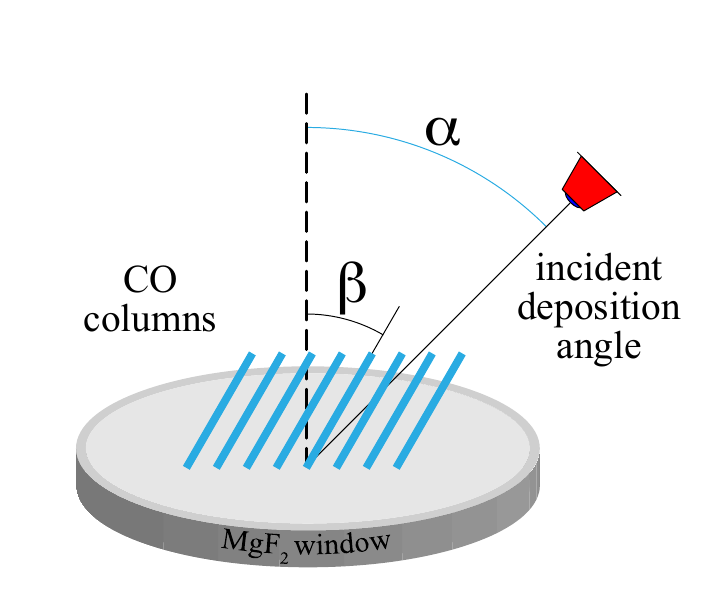}
	\caption{Schematic representation of the deposition angle, $\alpha$ in degrees, and the  tilt angle of the columns, $\beta$.}
	\label{Fig_columnas}
\end{figure}

This tendency is strikingly similar to the CO photodesorption rate versus the deposition angle, see Fig.~\ref{Fig_pico70}, suggesting that the CO photodesorption rate in these experiments depends mainly on the apparent ice surface area. It is thus worth to gain a better understanding of ice surface morphology as a function of the deposition angle. Fortunately, as mentioned above, there is ample literature that discusses this process for a variety of films of different atomic or molecular composition. At the low deposition temperature of 11 K, diffusion of CO molecules is limited as our TPD results suggest, and the CO ice will grow following a ballistic deposition. This growth proceeds by "hit-and-stick" or ballistic deposition and the molecules stay close to their landing site position. At normal incidence, a dense film of uniform thickness grows. But for increasing incident deposition angle $\alpha$, the ice thickness is not uniform because the elevated spots that occasionally appear are not smoothed out as in the normal incidence case. Instead, they intercept subsequent impinging molecules and shadow the lower areas of the ice film. This process is known as self-shadowing and produces a porous film made of a continuous reticulated structure, and for larger angles, a uniform array of tilted nanocolumns with angle $\beta$ with respect to the cold substrate, see Fig.~\ref{Fig_columnas}. The volume of large voids increases substantially at $\alpha$ = 60$^{\circ}$ and onward, and the columns become distinguishable. The estimated maximum in the surface area near 70$^{\circ}$ was explained by the competition of increase and decrease of the surface area due to the columnar formation and due to increase of the columnar spacing and thickness \citep{Suzuki2001}. Quoting these authors, "it is remarkable that the dependence of the effective surface area on $\alpha$ is quite similar to that of the photocatalytic properties of TiO$_2$ thin films prepared by dynamic oblique deposition". Indeed, the most efficient photocatalysis occurred at $\alpha$ = 70$^{\circ}$ in \cite{Suzuki2001}. 

\begin{figure}
	\centering
	\includegraphics[width=\hsize]{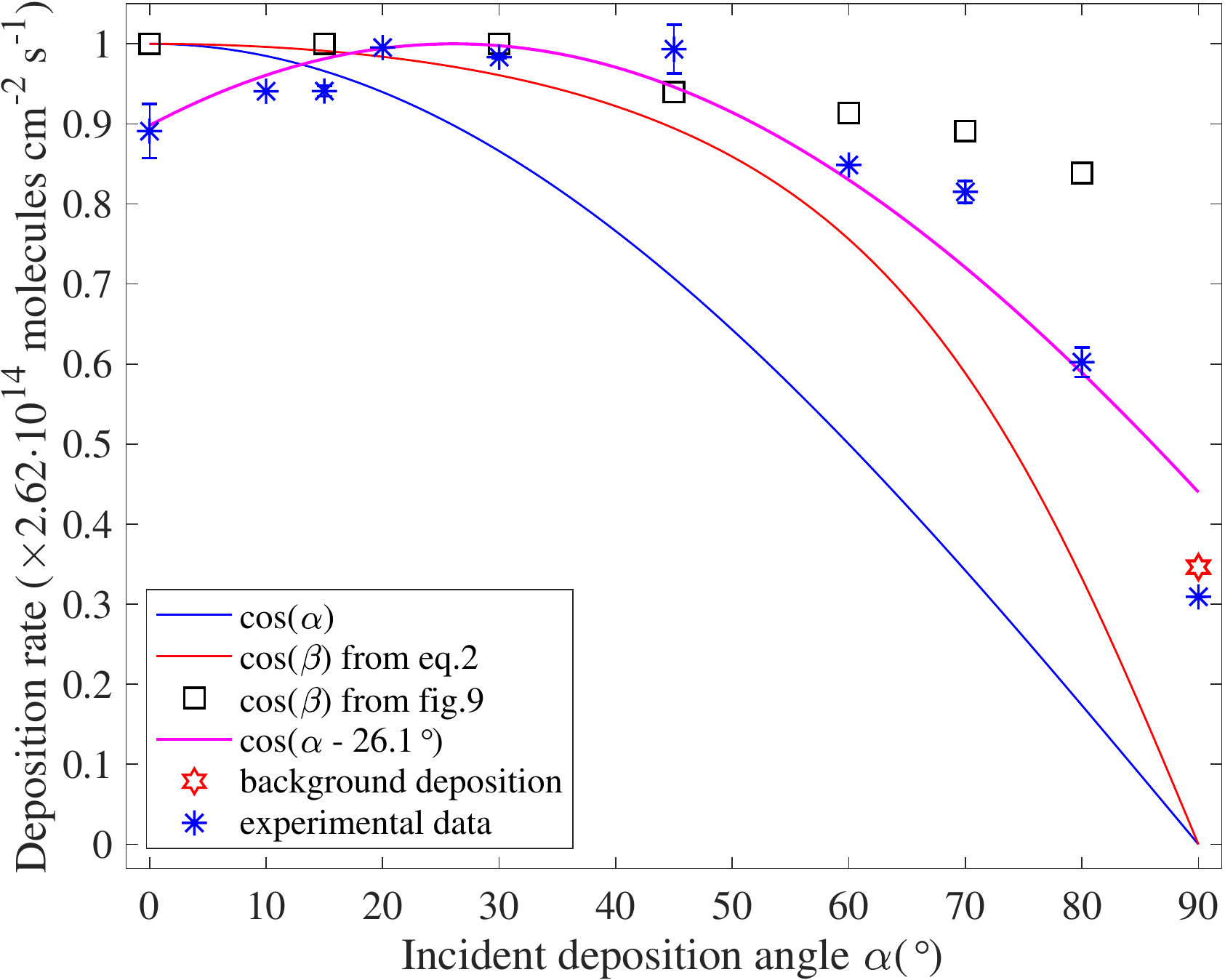}
	\caption{Deposition rate as a function of the incident angle $\alpha$ of deposition.}
	\label{Fig_dep_rate}
\end{figure}

In our CO ice irradiation experiments, the CO photodesorption rate with increasing deposition angle $\alpha$ also displays a maximum around 70$^{\circ}$, see Fig.~\ref{Fig_pico70}, and resembles, in addition to the aforementioned adsorption of gas on an amorphous water ice surface \citep{Dohnalek2003}, the i) modeled surface area at different angles in Fig.~10 of \cite{Suzuki2001}, and corresponds best to particles in cubic cells of length $d$ = 3--4 units, and ii) the specific surface area of TiO$_2$ and TiC grown at 100 and 77 K, respectively, at different angles  \citep{Suzuki2001}. In conclusion, in our experiments the photodesorption rate of CO ice measured at different deposition angles is not directly linked to the CO ice density. Instead, it follows the variations in the surface area of the ice versus deposition angle. This tendency might have been foreseen, since we mentioned that photodesorption of CO occurs at the ice surface when photons are absorbed in the top-most MLs \citep{Bertin2012}. In particular, for ices deposited between 7 K and 14 K, the UV photons that contribute to a photodesorption event are those absorbed in the top 5$\pm1$ ML \citep{MunozCaro2010,MunozCaro2016,Fayolle2011,Chen_2014}.

The rise of the photodesorption rate above 60$^{\circ}$ coincides with the appearance of tilted nanocolumns in films of different compositions, where $\beta$ represents the angle between the column and the cold surface, thus increasing the effective surface area of the ice with respect to normal angle deposition. The unshadowed areas of these columns are exposed to the UV flux of the MDHL and lead to a more efficient photodesorption than the ice deposited at normal incidence, which indicates a column thickness more than 5 ML or about 1.6 nm, a value compatible with the column thickness measured by imaging of metallic films deposited at oblique angles, see \cite{BARRANCO2016}.   

\begin{figure}
	\centering
	\includegraphics[width=\hsize]{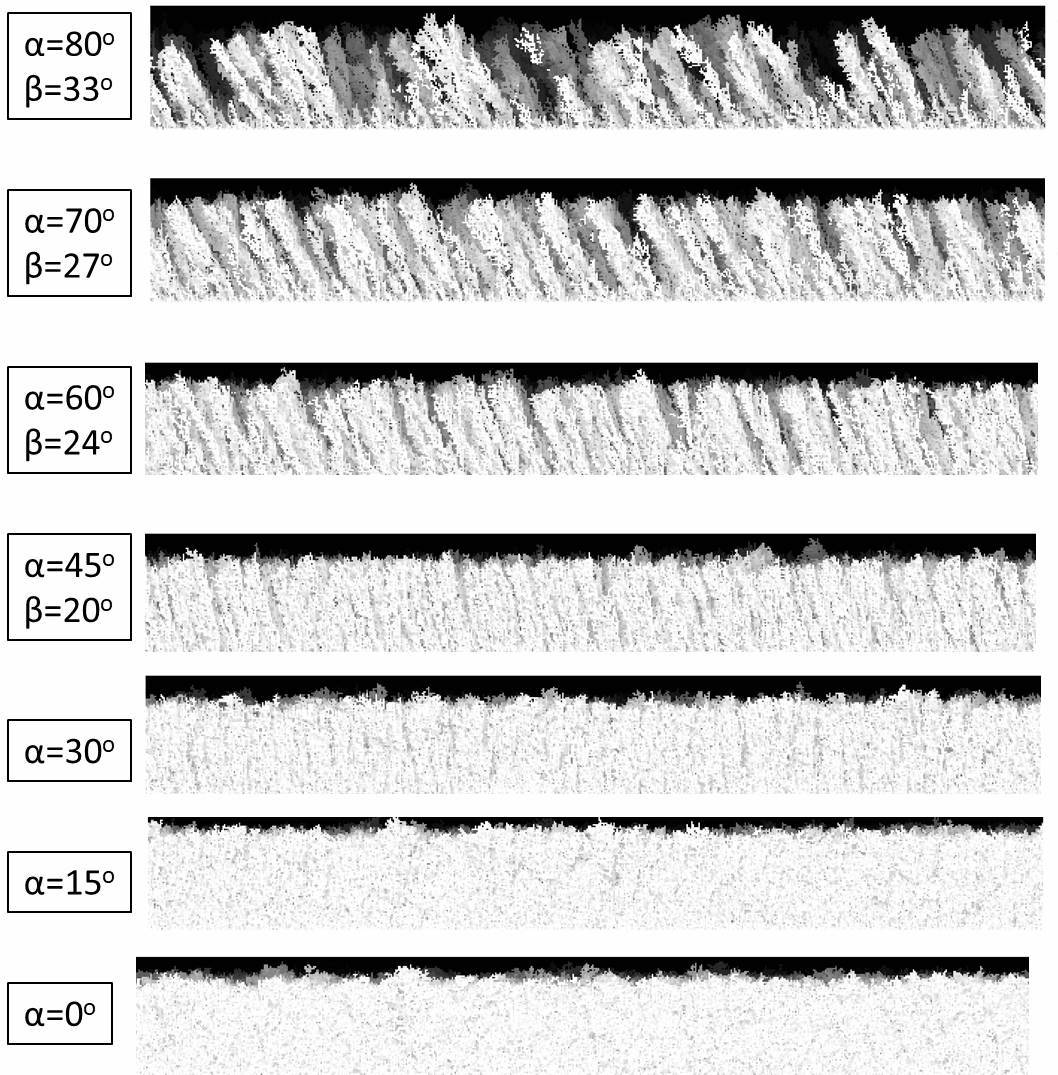}
	\caption[]{Two-dimensional section through the 80 nm deposition abtained from the surface trapping in oblique nanostructured growths (STRONG) software \citep{Alvarez2014}.}
	\label{Fig_columnsbeta}
\end{figure}

Another indication of the impact of deposition angle on ice morphology is provided by the ice deposition rate corresponding to different angles, see Fig.~\ref{Fig_dep_rate}. The blue asterisks are the deposition rate at different angles, while the red star represents the background deposition. The excess in the deposition rate with respect to $cos(\alpha$) is explained by the formation of elevated areas leading to a columnar morphology at large angles, as explained above, which traps more molecules during the oblique flow deposition. A better fit of the data, red solid line, is delivered by the $cos(\beta$) where
\begin{equation}
tan(\alpha) = 2~tan(\beta)
\end{equation}
is a commonly used rule \cite[e.g.,][]{BARRANCO2016}, since $\beta$, the tilt angle of the columns, is closer to the real ice surface inclination than $\alpha$. 
Fig.~\ref{Fig_columnsbeta} shows images of film depositions that we simulated using the surface trapping in oblique nanostructured growths (STRONG) software \citep{Alvarez2014}. In this Monte Carlo ballistic model, the species follow straight trajectories until they are only 3--4 $\AA$ away from the surface and may bend their trajectory due to short-range interactions.  As input, we provided a sample surface of 600$\times$600 nm, a film thickness of 80 nm that matches the ice column density in our experiments, angular flux aperture of 6$^{\circ}$ at the substrate position, and a surface trapping probability of 1, which should be close to the value of CO ice at 10 K \citep{Cazaux_2017}, and therefore, surface shadowing is expected to dominate over thermally activated mobility. Between 0$^{\circ}$ and 30$^{\circ}$ the columns are not defined yet. The values of $\beta$ that correspond to 45$^{\circ}$, 60$^{\circ}$, 70$^{\circ}$, and 80$^{\circ}$ are 20$^{\circ}$, 24$^{\circ}$, 27$^{\circ}$, and 33$^{\circ}$, respectively. Black empty squares in Fig.~\ref{Fig_dep_rate} represent $cos(\beta$) for these values of $\beta$.

Finally, the best fit to our experimental data in Fig.~\ref{Fig_dep_rate} is provided by $cos(\alpha$ - 26.1$^{\circ})$. A similar expression is found for $\beta$ in \cite{TAIT1993}, where $\beta = \alpha - 16^{\circ}$ for large values of the incident angle $\alpha$. The precise value of $\beta$ depends on the material and, to our knowledge, it has not been measured for any ice films \citep{Zhu2012,Zhao2012}.

CO gas molecules in dense clouds are expected to impinge at any angle on the cold dust surfaces near 10 K. Diffusion will be low at dust temperatures of 10 K and increase at temperatures near the thermal CO desorption, as discussed in Sect.~\ref{Simulations}. Dust grains, where CO molecules accrete in dense clouds and young stellar objects, are already covered by a water-dominated ice mantle. Unlike the flat substrate windows used for ice deposition in most experiments, including our own, dust grain surfaces are likely not smooth, leading to large effective ice surface areas. 
Indeed, a canonical 0.01 $\mu$m-thick water-rich ice mantle, on which CO ice accretes, probably displays an effective surface area and topology still similar to the bare grain.   
The photodesorption rate in realistic ice mantles may, therefore, be closer to the maximum value we obtained at 70$^{\circ}$ incident angle of deposition in Fig.~\ref{Fig_pico70}.

\section*{Acknowledgements}
	\label{acknowledgements}
	We are grateful to M. A. Satorre, from Universitat Polit\`ecnica de Val\`encia in Alcoi, for discussions. 
	The Spanish Ministry of Science, Innovation and Universities supported this research under grant number AYA2017-85322-R (AEI/FEDER, UE), MDM-2017-0737 Unidad de Excelencia "Mar\'{i}a de Maeztu"-- Centro de Astrobiolog\'{i}a (INTA-CSIC), Retos Investigaci\'{o}n [BIA2016-77992-R (AEI/FEDER, UE)], and "Explora Ciencia y Explora Tecnolog\'{i}a" [AYA2017-91062-EXP].
	We also benefited from financial support by MOST grants in Taiwan: MOST 103- 2112-M-008-025-MY3.




\bibliographystyle{mnras}
\bibliography{biblio} 
\label{biblio}

\end{document}